\documentclass[aps,prc,preprint,tightenlines,groupedaddress,showpacs]{revtex4}
\usepackage{graphicx}
\begin{document}

\preprint{DFF 423--03--05}

\title{\bf Isospin fluctuations in spinodal decomposition
}
\author{M. Colonna}
\affiliation{{\small\it Laboratori Nazionali del Sud, 
        Via S. Sofia 44, I-95123, Catania, Italy}}
\author{F. Matera}
\email{matera@fi.infn.it}
\affiliation{{\small\it Dipartimento di Fisica, Universit\`a degli 
Studi di Firenze,}}
\affiliation{
{\small\it Istituto Nazionale di Fisica Nucleare, Sezione di
Firenze,}\\
{\small\it Via G. Sansone 1, I-50019, Sesto F.no (Firenze), Italy}}


\begin{abstract}
We study  the isospin dynamics in fragment formation
within the framework of an analytical model based on the spinodal
decomposition scenario.  We calculate the probability to obtain fragments
with given charge and neutron number, focussing on the derivation
of the width of the isotopic distributions.  Within our approach   
this is  determined by the dispersion of $N/Z$ among 
the leading unstable modes, due 
to the competition between Coulomb and symmetry energy 
effects, and by isovector-like fluctuations present in the matter that
undergoes the spinodal decomposition.   
Hence the widths exhibit a clear dependence on the properties 
of the Equation of State. By comparing two systems with different 
values of the charge asymmetry we find that
the isotopic distributions reproduce an isoscaling relationship.  
\end{abstract}

\maketitle

\section{\label{}Introduction}
In the last years a widespread attention has been devoted to the role 
played by the isospin degree of freedom in the heavy--ion reaction 
physics. The interest on this subject is twofold: the knowledge 
of the symmetry term in the Equation of State (EOS) of  
asymmetric nuclear matter, which is a fundamental ingredient in 
astrophysical investigations \cite{Lat01}, and the 
thermostatistical properties  both at equilibrium and out of equilibrium 
of systems with two strongly interacting components 
\cite{Mue95,Bao97,Bar98,Lar99,Bot01,Bar01,Bar02}. Both the interests concern 
systems faraway from the physical conditions of ordinary nuclear matter.  
\par  
Thanks to the availability of high--performance $4\pi$--detectors 
for the investigations of heavy--ion collisions at intermediate  energy 
\cite{Xu00,Tsa01,Xu02,Ger04}, 
recent experimental results can provide new insights about isospin 
effects on the nuclear dynamics. In particular, for multifragmentation 
processes we can obtain information about highly excited
two--component systems and their subsequent decomposition. 
Statistical models have been extensively applied to the description 
of experimental data, also for isospin observables \cite{Tan03},
and some conclusions have been drawn on the behavior of charge
asymmetric systems. These models, however, imply the  
achievement of the statistical equilibrium for the nuclear system. 
Then, it would be highly desireable to have some 
insight on the path followed by the system to attain equilibrium, 
if this occurs. Further, it would be of great advantage to envisage 
some observable, which preserves memory of the dynamical processes 
occurred during the fragmentation. 
\par
In this paper we present an analytical description of the
disassembly of excited nuclear systems 
formed during the collision of heavy ions, 
in terms of the occurrence of nuclear matter instabilities. Our  
approach accounts for the source of the density fluctuations occurring 
when the system enters the spinodal instability region 
of the density--temperature phase diagram, and describes  
the growth of the fluctuations with time until they cause the 
decomposition of the system. 
This approach is a generalization to include the isospin   
degree of freedom,  of the model developed in Refs. \cite{Mat00,Mat03} 
for symmetric nuclear matter basically. This gives rise to a 
substantial improvement of the model, with new valuable results.   
Such extension allows us to investigate separately fluctuations of the  
neutron and proton densities and their interplay. Following the procedure 
introduced in Ref. \cite{Mat00}, we identify the pattern of the domains 
containing correlated density fluctuations, with the fragmentation 
pattern, and can make predictions on 
the isotopic distributions of the fragments. Moreover, we include 
in the present treatment the Coulomb force according to the approach 
outlined in Ref. \cite{Fab98}. Its effects on the isotopic 
distributions turn out to be sizeable. 
\par
Our results essentially refer to the distributions of the fragments 
just after the early break--up of the system. 
So our approach can be considered 
complementary to dynamical model calculations 
based upon semiclassical kinetic equations 
for one--body phase--space density, (for a review on dynamical models 
see, e.g., Refs. \cite{Das01,Bor02,Cho04}),
as far as the description of the early fragmentation mechanism 
is concerned. The advantage here is that one can make significant 
predictions on observables of experimental 
interest on an analytical basis. This allows us to directly relate the 
results obtained to the EOS properties   
and the features of the spinodal mechanism.  
In our scheme the onset and the growth of the fluctuations about 
the mean phase--space density in unstable situations, are 
self--consinstently treated. The self--consistency condition is provided 
by the fluctuation--dissipation theorem. Whereas all the processes, which take 
place before the system enters the spinodal instability region 
and after the break--up, are beyond our approach. Therefore  
the mean values of density, temperature and 
asymmetry of the nuclear medium when the system starts to break up 
are taken from calculations performed within dynamical models. 

On the other side, a dynamical model, which appropriately 
incorporates the effects of the 
fluctuations, might give a detailed description of the whole history 
of a collision between heavy ions. 
Therefore, it can be of interest to compare the results of our approach  
with those obtained by numerical solutions of microscopic 
transport equations, also to connect the results of the simulations to 
what is expected in a pure spinodal decomposition scenario. 
The comparison will be done  with the isotopic 
distributions for the primary fragments, calculated in 
the dynamical stochastic 
mean--field (SMF) approach of Refs. \cite{Bar02,Col98}. In particular,  
we will consider the ratio, for a given value of the proton number, 
between the isotope yields from two different reactions. 
This quantity represents a straightforward mean to compare isotopic 
distributions, since it is experimentally found to obey a simple 
relationship (isoscaling), as a function of the proton number and 
neutron number \cite{Xu00,Tsa01,Tan01,Tsa101}. 
We will also discuss the dependence of the isoscaling parameters
on the EOS considered. 
\par
In Sec. II we outline the extension of the formalism developed in 
Ref. \cite{Mat00} only for isoscalar density fluctuations,  
to include the isospin degree of freedom. In 
Sec. III we discuss the results of our calculations and their comparison 
with the calculations performed in Ref. \cite{Liu04} within the 
SMF approach. Finally, in Sec. IV a brief summary and conclusions are given. 
 
\section{\label{AA}Formalism} 
\subsection{Time evolution of density fluctuations}
We study the density fluctuations by introducing a self--consistent 
stochastic field acting on the constituents of the system. 
The time evolution of the fluctuations is described by a kinetic equation, 
within  a linear approximation for the stochastic field.  
The growth of fluctuations is
essentially dominated by the unstable mean field. Thus
we focus our attention on the behavior of the mean field and
neglect the collision term in the kinetic equation.
Collisions would mainly add a damping to the growth rate
of the fluctuations and should not change the main results of our calculations,
at least at a qualitative level. 
\par
The additional stochastic mean field, which we assume having 
a vanishing mean, will induce 
fluctuations of the proton and neutron densities, 
$\delta\varrho_i({\bf r},t)$, with respect to their uniform mean values 
$\varrho_i$ (~$i=1,2$ for protons and neutrons respectively~). 
We assume that at the time $t=0$, given density fluctuations 
$\delta\varrho_i({\bf r},t=0)$ are present in the system. 
The equations for the Fourier coefficients of $\delta\varrho_{i}({\bf r},t)$  
for $t>0$ are given by a generalization of the equation for the 
isoscalar density fluctuations of Ref. \cite{Mat00,Lalime}. They read 
\begin{eqnarray}
\delta\varrho_i({\bf k},t)=&& 
\delta\varrho_i({\bf k},t=0)-\Sigma_{j,l}\,\delta\varrho_l({\bf k},t=0)
D_{j,l}^{-1}(k,\omega=0 )\,
\int_{0}^{t}D_{i,j}(k,t-t^\prime)\,dt^\prime
\nonumber\\
&&+\Sigma_j\int_{0}^{t}D_{i,j}(k,t,t^\prime)dW_j({\bf k},t^\prime)\,. 
\label{wiener}
\end{eqnarray}
where the $2\times 2$ matrix in the isospin space, 
$D_{i,j}(k,t-t^\prime)$, is the density--density 
response function and $D_{i,j}(k,\omega)$ its time Fourier transform. For 
symmetry reasons the response function and its Fourier transform depend 
only on the magnitude of the wave vector. In the last integral 
$dW_j({\bf k},t^\prime)$ gives the contribution of the $j$--component of  
the stochastic field in the interval $dt^\prime$. Since the stochastic 
field is real  $W_i^{*}({\bf k},t)=W_i({-\bf k},t)$. The real and imaginary 
parts of the Fourier coefficients $W_{i}({\bf k},t)$ are 
indipendent components of a multivariate stochastic process \cite{Gard}, 
with
\begin{equation}
<\int_{0}^t\,dW_i({\bf k},t^\prime)\int_{0}^t\,dW_j(-{\bf k},t^{\prime\prime})>
=\int_{0}^tdt^{\prime}dt^{\prime\prime}\,
B_{i,j}({\bf k},t^\prime,t^{\prime\prime})\,
\label{wiener1}
\end{equation}
defining the correlator for the stochastic field. Angular brackets denote 
ensemble averaging.
\par
In the mean--field approximation the response function obeys the 
following set of equations 
\begin{equation}
D_{i,j}(k,\omega)=\,D^{(0)}_i(k,\omega)\delta_{i,j}+\Sigma_lD^{(0)}_i(k,\omega)
{\cal A}_{i,l}(k)D_{l,j}(k,\omega)\,
\label{resp}
\end{equation}
where $D^{(0)}_i(k,\omega)$ is the non--interacting particle--hole 
propagator and ${\cal A}_{i,l}(k)$ is the Fourier transform of the 
nucleon--nucleon effective interaction.  
\par
In Ref. \cite{Mat00} it has been shown that, in the case 
of isoscalar fluctuations in symmetric  
nuclear matter, a white--noise hypothesis for the stochastic field  
can be retained for values of temperature and density
sufficiently close to the borders of the spinodal
region. In such situations the imaginary part of the response function 
displays a sharp peak dominating the particle--hole background 
at a value of $\omega\ll kv_F$. This is due to the occurrence of a pole on the
imaginary axis of $\omega$, that corresponds to isoscalar fluctuations, 
at a distance from the origin that is much smaller
than the values of $kv_F$. The position of this pole determines
the time scale characteristic of the response function. 

However, when one wants to investigate the properties of 
neutron and proton distributions, as we do in the present study, one should
consider also the effects due to the isovector fluctuations. Even though 
isoscalar modes are the dominant ones, since they are unstable, isovector
fluctuations contribute to the width of the isotopic distributions of the
fragments formed in the spinodal decomposition process. 
In asymmetric nuclear matter isovector and isoscalar fluctuations are coupled. 
However one can still distinguish oscillations with neutrons and protons
moving in phase (isoscalar-like) or out of phase (isovector-like).
Let us first concentrate on the properties of the isoscalar-like modes. 

\subsubsection {Isoscalar-like fluctuations}

The position of the pole $\omega=i\Gamma_k$ 
for the unstable isoscalar-like mode 
is given by 
the imaginary root of the equation
\begin{equation}
{\rm det}|\delta_{i,j}-D^{(0)}_i(k,\omega){\cal A}_{i,j}(k)|=0\,.
\label{rate}
\end{equation}
The quantity $\Gamma_k$ is the damping or growth rate (depending on its 
sign) of the density fluctuations. In evaluating it, we use the 
expression of $D^{(0)}_i(k,\omega)$ for $\omega\ll kv_F$ \cite{Mat00} 
\[
D_{i}^{(0)}(k,\omega)\simeq -\frac{\partial \varrho_i}{\partial \tilde\mu_i}
-i\frac{1}{2\pi}m^2F(\beta \tilde\mu_i)\frac{\omega}{k}\, ,
\] 
where the effective chemical potential $\tilde\mu_i$ of neutrons or 
protons is measured with respect to the uniform mean field 
$U_i(\varrho_1,\varrho_2)$ of the unperturbed initial state and 
$F(\beta \tilde\mu_i)$ is the function 
\[F(\beta \tilde\mu_i)=\,\frac{1}{e^{-\beta \tilde\mu_i}+1}\,,\]
with $\beta =1/T$ being the inverse temperature (we use units
such that $\hbar=~c=~k_B=1$).
\par
Substituting into Eq. (\ref{wiener}) the response function 
$D_{i,j}(k,t-t^\prime)$ calculated with these approximations, the equation 
for the fluctuations $\delta\varrho_i({\bf k},t)$ becomes 
\begin{eqnarray}
\delta\varrho_i({\bf k},t)=&&\delta\varrho_i({\bf k},t=0)+
\Sigma_{j,l}C_{i,l}(k)D_{l,j}^{-1}(k,\omega=0)
\delta\varrho_j({\bf k},t=0)\frac{1}{\Gamma_k}(e^{\Gamma{_kt}}-1)
\nonumber\\
&&+\Sigma_j\,C_{i,j}(k)e^{\Gamma{_kt}}\int_0^te^{-\Gamma{_kt^\prime}}
\,dW_j({\bf k},t^\prime)\, ,
\label{ornul}
\end{eqnarray}
where $C_{i,j}(k)$ are the residues, times $(-i)$, of the components of the 
response function at the pole $\omega=i\Gamma_k$. They have the 
relevant property 
\begin{equation}
{\rm det}|C_{i,j}(k)|=0\,.
\label{det}
\end{equation}
The explicit expression of the inverse of the response function 
for $\omega=0$ is 
\[D_{i,j}^{-1}(k,\omega=0)=-\Big[\frac{\partial \tilde\mu_j}
{\partial \varrho_i}+{\cal A}_{i,j}(k)\Big]\,.\]
\par
For isoscalar-like fluctuations $W_j({\bf k},t^\prime)$ represents a 
Gaussian white noise \cite{Mat00}. 
The probability distribution of density fluctuations, 
$P[\delta\varrho_i({\bf k},t)]$, is given by a product of Gaussian 
distributions. Each single factor 
corresponds to the stochastic process of Eq. (\ref{ornul}) 
for a given wave number $k$ \cite{Mat00,Mat03}, with 
the covariance matrix 
\begin{equation}
\sigma^2_{i,j}(k,t)=\Sigma_{l,m}C_{i,l}(k)B_{l,m}({\bf k},t)C_{m,j}(k)
\frac{1}{2\Gamma_k}\Big(e^{2\Gamma_kt}-1\Big)\,.
\label{variance0}
\end{equation}
For simplicity, we have assumed that the initial fluctuations are 
negligible $\sigma^2_{i,j}(k,t)\simeq 0$. Whenever it is necessary, a 
nonvanishing covariance can be easily introduced. 
\par
The probability distribution $P[\delta\varrho_i({\bf k},t)]$ is completely 
determined once the covariance matrix $\sigma^2_{i,j}(k,t)$  
is known. According to the 
procedure usually followed when treating instabilities by exploiting  
the fluctuation--dissipation theorem, see e.g. Refs. \cite{Gunt83,Hoff95}, 
we determine the coefficients $B_{i,j}({\bf k},t)$ 
as functions of $\varrho_1$, $\varrho_2$ and $T$ for the
system at equilibrium, then we extend the
expressions so found to non--equilibrium cases. 
Since the relevant
values of the wave vector $k$ turn out to be such that
the quantity  $kv_F$ is of the same order of magnitude as $T$,
the limit $\omega/kv_F\ll 1$ also implies  $\omega/T\ll 1$. In such 
case, the classical limit $\omega/T\ll 1$ (or $|\Gamma_k(t)|/T\ll 1$) 
can be taken when evaluating both sides of the 
fluctuation--dissipation relation. Then, we get 
\begin{equation}
\frac{\partial}{\partial t}
<\delta\varrho_i({\bf k},t)\delta\varrho_j({-\bf k},t^\prime)>=
-TD_{i,j}(k,t-t^\prime)\,.
\label{fdt}
\end{equation}
The equation for the equilibrium fluctuations can 
be obtained from Eq. (\ref{wiener}) by shifting the initial time 
$t=0$ to $-\infty$. 
By exploiting Eq. (\ref{fdt}) we can obtain 
the following relation between the coefficients 
$B_{i,j}({\bf k},t)$ and the functions $C_{i,j}(k)$:
\begin{equation}
\Sigma_{l,m}C_{i,l}(k)B_{l,m}({\bf k},t)C_{m,j}(k)=\,
-2TC_{i,j}(k)\,.
\label{fdt1}
\end{equation}
From this equation we can see that $B_{i,j}$ are constant and 
depend only on the magnitude 
$k$ of the wave vector, as it is expected for symmetry reasons. 
Following Refs. \cite{Gunt83,Hoff95} (see also the discussion in 
Ref. \cite{Mat00} on this point) we assume that the relation (\ref{fdt1}) 
is valid also in instability situations. In such a way, the covariance 
matrix (\ref{variance0}) acquires the form 
\begin{equation}
\sigma^2_{i,j}(k,t)=-TC_{i,j}(k)
\frac{1}{\Gamma_k}\Big(e^{2\Gamma_kt}-1\Big)\,,
\label{variance}
\end{equation}
and is completely determined both for stable and  
unstable situations.
We notice that, for the isoscalar-like mode, $\sigma^2_{1,2}(k) = 
\sigma^2_{2,1}(k)$ is positive. In fact proton and neutron 
densities oscillate in phase, although with different amplitudes 
in general. However, the ratio between amplitudes,
$\sigma^2_{1,1}(k)/ \sigma^2_{1,2}(k)$,  is found to be larger 
than the initial proton to neutron ratio, thus leading to 
the formation of more symmetric fragments, the so-called isospin 
distillation effect \cite{Bar98}.

\subsubsection {Isovector-like fluctuations}

Now we turn to consider the isovector-like modes. 
In this case the frequency of the modes, $\omega^{iv}_k$ is real, 
i.e. we have stationary oscillations.  The position of the pole is given
by the other solution of Eq. (\ref{det}). However, 
we add a small negative imaginary part $-\Gamma^{iv}_k$ 
to the position of the pole, taking into account that here 
we are neglecting nucleon-nucleon collisions and finite size effects. 
Correspondingly the imaginary part of the response function acquires the  
width $\Gamma^{iv}_k$. \par
The contribution of isovector-like fluctuations to 
the covariance matrix $\sigma_{i,j}^2(k,t)$ can be written 
as it follows:
\begin{eqnarray}
\sigma^2_{i,j}(k,t)&&=4\,\Sigma_{l,m}C_{i,l}^{iv}(k)C_{m,j}^{iv}(k)
e^{-2\Gamma^{iv}_k t}
\nonumber\\
&&\times\int_0^tdt_1dt_1^{\prime}\Big[e^{\Gamma^{iv}_k(t_1+t_1^{\prime})}
B_{l,m}^{iv}({\bf k},t_1,t_1^{\prime})
\sin\big(2\omega^{iv}_k(t-t_1)\big)
\sin\big(2\omega^{iv}_k(t-t_1^{\prime})\big)\Big]\,,
\label{fluc_iv}
\end{eqnarray}
where $C_{i,j}^{iv}(k)$ are the residues at the pole and 
$B_{l,m}^{iv}({\bf k},t_1,t_1^{\prime})$ denote the contributions from the 
isovector--like fluctuations to the stochastic field. 
\par
To determine the amplitude of the stochastic field we essentially 
follow again the derivation presented above. 
By exploiting the fluctuation-dissipation theorem, now 
in the limit $\omega/T>>1$
(since the frequency of the isovector vibrations is rather large with respect
 to the relevant values of $T$), we obtain for values of 
$\omega$ close to the pole the relation: 
\begin{equation}
\Sigma_{l,m}C_{i,l}^{iv}(k)B_{l,m}^{iv}({\bf k},\omega)C_{m,j}^{iv}(k)=\,
2\Gamma^{iv}_k C_{i,j}^{iv}(k)\Big(\,\frac{2(\Gamma^{iv}_k)^2}{(\omega-
\omega^{iv}_k)^2+(\Gamma^{iv}_k)^2}\Big)\,, 
\label{fdt1_iv}
\end{equation}
where we have added a Lorentzian factor to the right hand side in order to 
restrict to a small region about $\omega^{iv}_k$ the contribution 
from the isovector--like pole to the time Fourier transform of 
$B_{l,m}^{iv}({\bf k},\omega)$. In this way 
the correlator $B_{l,m}^{iv}({\bf k},t_1-t_1^{\prime})$ 
for the stochastic field results to be proportional to 
$e^{-\Gamma^{iv}_k|t_1-t_1^{\prime}|}$. This means that the isovector--like 
stochastic field is given by a coloured noise, at variance with the 
isoscalar case. 
\par
Substituting the time Fourier transform of Eq. (\ref{fdt1_iv}) into  
Eq. (\ref{fluc_iv}), 
and retaining only the leading term of the expansion in powers of 
$(\Gamma^{iv}_k/\omega^{iv}_k)$, we obtain for the covariance matrix 
the expression 
\begin{equation}
\sigma^2_{i,j}(k,t)= C_{i,j}^{iv}(k)\Big(1-e^{-2\Gamma^{iv}_kt} 
-2\Gamma^{iv}_kt\,e^{-2\Gamma^{iv}_kt}\Big)+
O\big((\frac{\Gamma^{iv}_k}{\omega^{iv}_k})^2\big)\,,
\label{var_isov1}
\end{equation}
whose asymptotic value is given by 
\begin{equation}
\sigma^2_{i,j}(k) = C_{i,j}^{iv}(k)\,.
\label{var_isov}
\end{equation}
We notice that, for isovector-like fluctuations, 
$\sigma^2_{1,2}(k) = \sigma^2_{2,1}(k)$ is negative. Indeed neutron and proton 
densities oscillate out of phase. \par
The covariance matrix of Eq. (\ref{var_isov}) refers to equilibrium 
fluctuations at given values of density and charge asymmetry. It can be 
directly obtained by means of the fluctuation--dissipation relation in the 
case of a purely real pole (~$\Gamma^{iv}_k\rightarrow 0$~).\par
We finally remark that the covariance matrix of Eq. (\ref{var_isov}) is 
obtained in the limit $T\rightarrow 0$ and, in addition, it does not depend 
on the width $\Gamma^{iv}_k$ of the isovector--like resonance. This implies 
that the density fluctuations of isovector--like nature, we are considering,  
have a quantum origin. 

\subsection{Size distributions}

Now we describe the procedure to determine the distribution for the size 
of the correlation domains. 
We closely follow the derivation given in 
Ref. \cite{Mat00} for isoscalar density fluctuations, and we limit 
ourselves to outline the steps relevant to the present more general 
treatment. We distinguish the fluctuations of the proton density from 
those of the neutron density. 

\subsubsection {Correlation lengths}
 
The probability distribution for the sizes of the domains 
where the fluctuations are correlated, $b_1$ and $b_2$ for protons and 
neutrons respectively, can be obtained by means of the functional integral 
\begin{eqnarray}
P(b_1,b_2,t)=&&\,\int d[\delta\varrho_i({\bf r},t)]\,\delta\bigg(b_1
-\int d{\bf r}d{\bf r}^\prime\delta\varrho_1({\bf r},t)f_1({\bf r})
\delta\varrho_1({\bf r}^\prime,t)f_1({\bf r}^\prime)\bigg)
\nonumber
\\
&&\,\delta\bigg(b_2
-\int d{\bf r}d{\bf r}^\prime\delta\varrho_2({\bf r},t)f_2({\bf r})
\delta\varrho_2({\bf r}^\prime,t)f_2({\bf r}^\prime)
\bigg)P[\delta\varrho_i({\bf r},t)]\, ,
\label{pbi0}
\end{eqnarray}
where $P[\delta\varrho_i({\bf r},t)]$ is the probability distribution 
for the density fluctuations and $f_i({\bf r})$ are suitable weight 
functions. 
Moreover, we assume that the dynamical correlation lengths 
for proton and neutron density fluctuations, $<b_1>$ and $<b_2>$, coincide 
\begin{equation}
L(t)=\,\int\frac{d{\bf k}}{(2\pi)^3}\sigma^2_{1,1}(k,t)|f_1(k)|^2=\,
\int\frac{d{\bf k}}{(2\pi)^3}\sigma^2_{2,2}(k,t)|f_2(k)|^2\,,
\label{corrl}
\end{equation}
where $f_i(k)$ are the Fourier transforms of the weight functions. 
In this way we assume that, on average, neutrons and protons are correlated
within the same domain. 
We will see in the following how this can be related
to the average isospin distillation effect in the formation of fragments.  
\par
Following the procedure used in Ref. \cite{Mat00} 
we obtain for the probability distribution $P(b_1,b_2,t)$ the equation 
\begin{eqnarray}
P(b_1,b_2,t)=&&\frac{1}{2\pi}\frac{1}{L(t)}\,\frac{1}
{[b_1+b_2]}\frac{1}
{\sqrt{\gamma(t)}}{\rm exp}\bigg(-\frac{[b_1+b_2]}
{4L(t)}\bigg)
\nonumber\\
&&\times{\rm exp}\bigg(-\frac{1}{4L(t)\gamma(t)}\frac{
[b_1-b_2]^2}{[b_1+b_2]}
\bigg)\,,
\label{distrb}
\end{eqnarray}
\par
where the parameter $\gamma(t)$ is given by  
\begin{equation}
\gamma(t)=1-\frac{\int d{\bf k}\sigma^2_{1,2}(k,t)|f_1(k)|^2
\int d{\bf k}\sigma^2_{1,2}(k,t)|f_2(k)|^2}
{\int d{\bf k}\sigma^2_{1,1}(k,t)|f_1(k)|^2
\int d{\bf k}\sigma^2_{2,2}(k,t)|f_2(k)|^2}\,.
\label{gamma0}
\end{equation}
At variance with the case of isoscalar fluctuations, 
the distribution $P(b_1,b_2,t)$ depends on the weight 
functions $f_i(k)$. These functions, to some extent, are 
arbitrary, the only requirement is 
that the integrals containing them should converge. 
For simplicity, we assume $|f_i(k)|^2=a_i|f(k)|^2$. 
For the functional form of $|f(k)|^2$ we choose the simplest one: 
$|f(k)|^2=1/k^2$. This choice is also supported by the fact that  
for equilibrium fluctuations 
the integral of the variance weighted with $1/k^2$  
gives the correct value of the correlation length \cite{Mat00}. 
In addition, we have found that for the physical situations 
considered in this paper, the value of the 
parameter $\gamma(t)$ to a large extent is insensible to the particular 
form of the weight function $|f(k)|^2$.
\par
From the probability distribution of the domain sizes we can obtain 
the distribution of the numbers of correlated protons $Z$ and neutrons $N$, 
assuming the correlation domains to be spherical. 
The relations between $Z$ and $b_1$, and $N$ and $b_2$ 
can be expressed as
$b_1=2r_{01}Z^{1/3}$ and $b_2=2r_{02}N^{1/3}$, where $r_{0i}$ is 
the mean interparticle spacing for nucleons of the $i$--species, 
calculated at the actual values of asymmetry and density (when fragments are
formed),  
that are different from asymmetry and density of the initial matter.  
The fact that the fragment size is related to the correlation length  
can be considered as a reasonable assumption in situations where
isoscalar-like modes are the dominant ones, as in fragmentation processes. 

So, since on average $b_1$ is equal to $b_2$, we obtain:
$r_{01}/r_{02}=(\rho_2/
\rho_1)^{1/3}=<N^{1/3}>/<Z^{1/3}>$, where $\rho_i$ are the densities
calculated at the time fragments are formed.
In this way the ratio $r_{01}/r_{02}$ can be 
related to the average asymmetry of the liquid (fragment) phase, 
obtained after the distillation process has occurred. 
One can consider, for instance, as average fragment 
asymmetry, values extracted from dynamical SMF simulations 
for primary fragments \cite{Bar02}. 

Then, the probability distribution of $Z$ protons and $N$ neutrons 
contained in a correlation domain, acquires the form  
\begin{eqnarray}
P(Z,N,t)=&&\frac{1}{9\pi}\frac{r_0}{L(t)}\,\frac{\lambda_1\lambda_2}
{[\lambda_1Z^{1/3}+\lambda_2N^{1/3}]}\frac{1}{(ZN)^{2/3}}\frac{1}
{\sqrt{\gamma(t)}}{\rm exp}\bigg(-\frac{r_0}{2L(t)}
[\lambda_1Z^{1/3}+\lambda_2N^{1/3}]\bigg)
\nonumber\\
&&\times{\rm exp}\bigg(-\frac{r_0}{2L(t)}\frac{1}{\gamma(t)}\frac{
[\lambda_1Z^{1/3}-\lambda_2N^{1/3}]^2}{[\lambda_1Z^{1/3}+\lambda_2N^{1/3}]}
\bigg)\,
\label{distrzn}
\end{eqnarray}
with $\lambda_i=r_{0i}/r_0$, where $r_0$ is the mean interparticle 
spacing for nucleons of both species. 

\subsubsection {Correlation volumes}

One may also assume that the size of fragments is directly 
related to a correlation volume $V$, instead of a correlation length. 
Equation (\ref{distrb}) can be rewritten for
the correlation volumes, just replacing $b_1$ and $b_2$ with 
$V_1$ and $V_2$. Then the probability distribution, after some algebra, 
reads: 
\begin{eqnarray}
P(Z,N,t)=&&\frac{1}{2\pi}\frac{1}{{\bar V}(t)}\,\frac{1}
{[\rho_2Z+\rho_1N]}\frac{1}
{\sqrt{\gamma(t)}}{\rm exp}\bigg(-\frac{1}{4{\bar V}(t)}
[Z/\rho_1+N/\rho_2]\bigg)
\nonumber\\
&&\times{\rm exp}\bigg(-\frac{1}{4{\bar V}(t)}\frac{1}{\gamma(t)}\frac{
[Z/\rho_1-N/\rho_2]^2}{[Z/\rho_1+N/\rho_2]}
\bigg)\,
\label{distrzn_volume}
\end{eqnarray}
where ${\bar V}$ is the average correlation volume for nucleons of both
species. 
For not too large asymmetries, this can be rewritten in the following form:
\begin{eqnarray}
P(Z,N,t)=&&\frac{1}{\pi A {\bar A}}\,\frac{1}
{\sqrt{\gamma(t)}}{\rm exp}\bigg(-\frac{A}{2{\bar A}}
\bigg)
\nonumber\\
&&\times{\rm exp}\bigg(-\frac{A}{2{\bar A}}\frac{1}{\gamma(t)}
\Big[\frac{N-Z}{A} - \alpha\Big]^2
\bigg)\,
\label{distrzn_stat}
\end{eqnarray}
where $\alpha= (\rho_2-\rho_1)/(\rho_2+\rho_1)$
represents the average asymmetry of fragments and ${\bar A}$ is
the average mass.  

\section{\label{BB}Results}

In our calculations we have adopted a schematic Skyrme--like effective 
interaction, that can be expressed as a sum of two terms 
\[
{\cal A}_{i,j}(k)= {\cal A}(k)+{\cal S}_{i,j}(k)\,.\]
For the symmetric term ${\cal A}(k)$ we use the finite--range effective 
interaction introduced in Ref. \cite{ColA94}: 
\begin{equation}
{\cal A}(k)=\Big(A\frac{1}{\varrho_{eq}}+(\sigma+1)\frac{B}
{\varrho_{eq}^{\sigma+1}}\varrho^{\sigma}\Big)e^{-c^2\,k^2/2}\,,
\label{inters}
\end{equation}
with $\varrho=\varrho_1+\varrho_2$ and 
\[
A=-356.8\,{\rm MeV},~~B=303.9\,{\rm MeV},~~\sigma=\,\frac{1}{6}\,.
\]
These values reproduce the binding energy
($15.75\,{\rm MeV}$) of symmetric nuclear matter at saturation 
($\varrho_{eq}=0.16\,{\rm fm}^{-3}$) and give an 
incompressibility modulus of $201\,{\rm MeV}$. 
The width of the Gaussian in Eq. (\ref{inters}) has been chosen in
order to reproduce the surface-energy term as prescribed in Ref. \cite{Mye66}.
\par
The isospin--dependent part, ${\cal S}_{i,j}(k)$, contains three different 
terms 
\begin{equation}
{\cal S}_{i,j}(k)=\frac{\partial^2{\cal E}_{symm}}{\partial\varrho_i
\partial\varrho_j}+\tau_i\tau_jDk^2+\frac{1+\tau_i}{2}V_C(k)\delta_{i,j}
\,,
\label{interv}
\end{equation}
with $\tau_1=1$ and $\tau_2=-1$. The double derivative 
of the potential part of the symmetry energy density, 
${\cal E}_{symm}$, is calculated in the unperturbed initial state. 
For the coefficient of the isovector surface term 
we use the value $D=40\,{\rm MeV\cdot fm}^5$ \cite{Bay71}.  
Concerning the Coulomb 
interaction, a mean--field exchange contribution 
\[V_C^{ex}=-\frac{1}{3}\Big(\frac{3}{\pi}\Big)^{1/3}e^2\varrho_1^{-2/3}
\] 
is added to the bare Coulomb force. 
\par
In order to stress the effects of the asymmetry of the nuclear medium, 
we will present results obtained with two different parametrizations 
of the symmetry energy: one with a stronger density dependence 
(~``superstiff'' asymmetry term~) and the other one with a weaker 
density dependence (~``soft'' asymmetry term~). 
In both cases the density dependence of the symmetry energy 
can be expressed by 
\[{\cal E}_{symm}(\varrho_1,\varrho_2)=S(\varrho)(\varrho_2-\varrho_1)^2
\,,\] 
with 
\begin{equation}
S(\varrho)=\frac{2d}{\varrho_{eq}^2}\frac{\varrho}{1+\varrho/\varrho_{eq}}
\,,
\label{stiff}
\end{equation}
where $d=19\,{\rm MeV}$ \cite{Bao00}, for the ``superstiff'' case, and 
\begin{equation}
S(\varrho)=d_1-d_2\varrho
\,,
\label{soft}
\end{equation}
where $d_1=240.9\,{\rm MeV\cdot fm}^3$ and $d_2=819.1\,{\rm MeV\cdot fm}^6$ 
\cite{ColA98}, for the ``soft'' case. 
\begin{figure}
\includegraphics{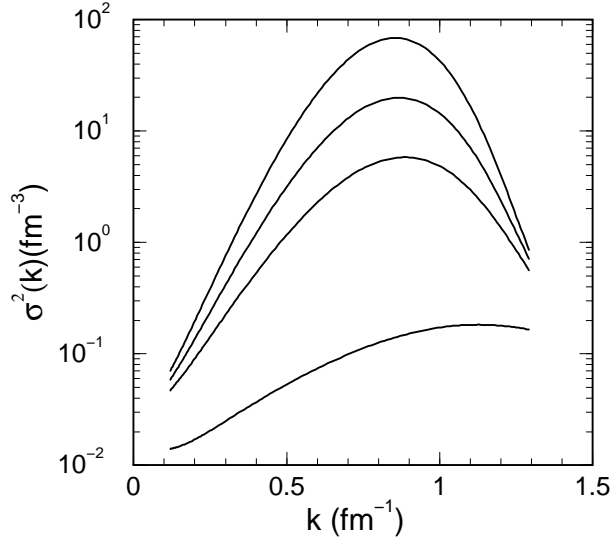}
\caption{\label{fig1}The variance for the unstable modes as a function 
of $k$ at four different times: from bottom to top $t=30,100,125,150
\,{\rm fm}/c$. The values of $\varrho$, ${\rm T}$, and $\alpha$ are 
$\varrho=0.3\varrho_{eq}$, ${\rm T}=4.5\,{\rm MeV}$, 
and $\alpha=0.2$. 
}
\end{figure}
\par
The inclusion of the Coulomb interaction presents sizeable  
effects on the stability conditions of nuclear matter. It gives rise 
to an overall decrease of the growth rate of density 
fluctuations with a corresponding contraction of the 
instability region in the ($\varrho,T$) phase diagram \cite{Fab98,
ColonnaPRL}. 
Moreover, it can be observed that, when the 
Coulomb force is included, the growth rate vanishes for  
sufficiently low values of the wave vector $k$ 
($k_{min}\simeq 0.2{\rm fm}^{-1}$) \cite{Fab98}. 
\par
In the integrals of Eqs. (\ref{corrl}) and (\ref{gamma0}), which 
determine the relevant parameters $L(t)$ and  $\gamma(t)$ 
for the distribution $P(Z,N,t)$, we 
consider only the contributions from the unstable modes. To this 
purpose, we put the weight function $f(k)$ equal to zero for $k$ 
larger than the value beyond which the rate $\Gamma_k$ becomes 
negative. However,
to evaluate the total value of the covariance matrix, 
we will consider the sum 
of the asymptotic value of the contribution due to isovector-like fluctuations,
Eq. (\ref{var_isov}) and the contribution due to the isoscalar-like modes, 
Eq. (\ref{variance0}), that grows exponentially. 

The variance for the unstable fluctuations of 
the isoscalar density, 
$\sigma^2(k) = \sigma^2_{1,1}(k) +  \sigma^2_{2,2}(k) + 2  \sigma^2_{1,2}(k)$,
is displayed in Fig.~\ref{fig1} at 
four different times. We only report the results obtained with the 
``superstiff'' symmetry term. For the isoscalar fluctuations 
the ``soft'' asymmetry term gives almost undistinguishable curves.  
The values chosen for the density $\varrho=0.3\varrho_{eq}$ and for 
the temperature $T=4.5\,{\rm MeV}$ are in the range expected for the 
multifragmentation process \cite{Cho04,Tam98}. For the asymmetry 
we choose a value of $\alpha=0.2$. Figure \ref{fig1} 
shows that the variance becomes a more and more peaked function 
about the most unstable mode with increasing time. It is worth noticing 
that the values of the variance of our calculations quite well 
compare with those obtained in Ref. \cite{Ayi96} within a different 
approach including the effects of the nucleon-nucleon collisions. 
This supports the 
suggestion that the development and the growth of the fluctuations 
are essentially determined by the instabilities of the mean field,
while the seeds are provided by the thermal agitation of the system. 
\par
We now turn to evaluate fragment isotopic distributions. 
In order to take into account that $Z$ and $N$ are discrete variables 
we express the probability of finding a correlation domain
containing $Z$ protons and $N$ neutrons, $Y(Z,N,t)$, through the integral
\begin{equation}
Y(Z,N,t)=\,\int _{Z-1}^{Z}dZ\int _{N-1}^{N}dN\,P(Z,N,t)\,. 
\label{probzn}
\end{equation}
For large $Z$ and $N$, $Y(Z,N,t)$ tends to coincide with $P(Z,N,t)$.  

We first consider Eq.(\ref{distrzn}) to calculate
the distribution $P(Z,N,t)$ and the probability $Y(Z,N,t)$. They are 
determined once the ratio $r_0/L(t)$ and the parameter 
$\gamma(t)$ have been calculated for given values of $\varrho$, $T$ 
and average asymmetry $\alpha$ of the system at the break--up. 
The length $L(t)$ characterizes the decrease 
of the correlation function with distance. The procedure to determine 
its value has been extensively discussed in Refs. \cite{Mat00,Mat03}. 
Here, we focus our attention on the calculation  
of the parameter $\gamma(t)$ 
characterizing the widths of the isotopic distributions. 
\par 
This can be evaluated by rewriting Eq. (\ref{gamma0}) with the assumptions 
about the weight functions introduced in Sec.~\ref{AA}: 
\begin{equation}
\gamma(t)=1-\frac{\int dk\sigma^2_{1,2}(k,t)|f(k)|^2
\int dk\sigma^2_{1,2}(k,t)|f(k)|^2}
{\int d k\sigma^2_{1,1}(k,t)|f(k)|^2
\int dk\sigma^2_{2,2}(k,t)|f(k)|^2}\,.
\label{gamma}
\end{equation}
\par
Since the magnitude of the isospin--distillation effect, i.e. the ratio
$\sigma_{1,2}^2(k)/\sigma_{1,1}^2(k) = \sigma_{2,2}^2(k)/\sigma_{1,2}^2(k)$,  
depends on the wave 
number $k$, even considering only the contribution of the isoscalar-like
modes to $\sigma_{i,j}^2(k)$, one obtains a non vanishing value of the width
$\gamma$.  Considering also the contribution of isovector-like fluctuations, 
the width $\gamma$ increases, as we will show in the following.  

For values of the asymmetry $\alpha$ of nuclear interest, the parameter 
$\gamma(t)$ turns out to be about $10^{-3}$ for both the considered 
asymmetry terms in the nucleon--nucleon interaction 
( ``soft'' and ``superstiff'' ). 
As a general trend, the parameter $\gamma(t)$ increases 
with increasing asymmetry and density of the decomposing system, and 
decreases with the time. 
\begin{figure}
\includegraphics{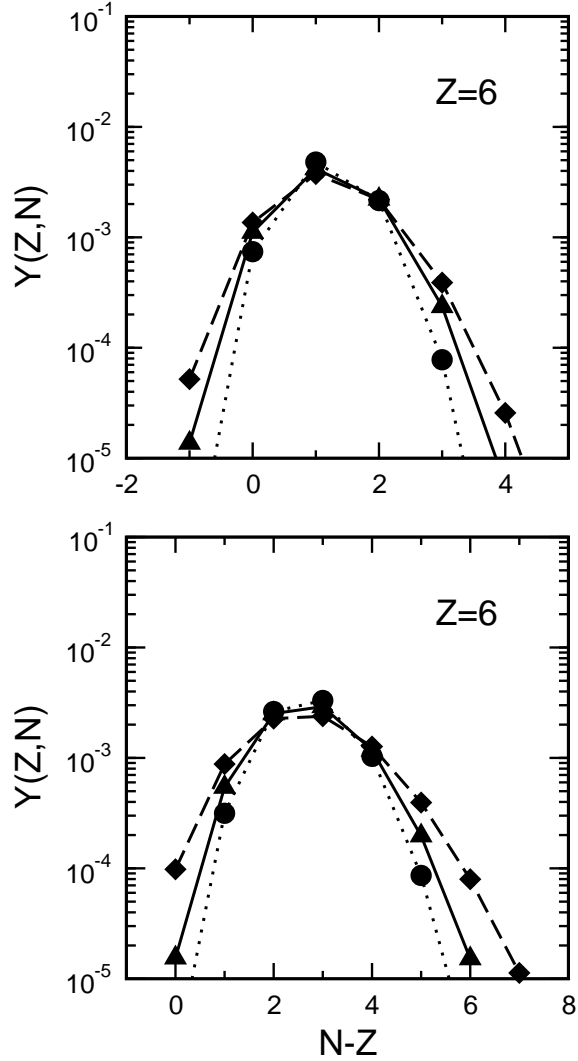}
\caption{\label{fig2}Calculated isotopic yields of $Z=6$--fragment with 
the ``superstiff'' symmetry term (diamonds) and the ``soft'' 
symmetry term (triangles).
The circles represent the results obtained neglecting the contribution
of isovector-like fluctuations in the ``soft'' case.  
The values of $\varrho$, ${\rm T}$, $L$, and $t$ are 
$\varrho=0.3\varrho_{eq}$, ${\rm T}=4.5\,{\rm MeV}$, 
$L=1.3\,r_0$, and $t=125\,{\rm fm}/c$. Top panel: $\alpha=0.1$,
the value of the parameter $\gamma(t)$ is $\gamma(t)=1.02\,10^{-3}$ 
for the ``superstiff'' symmetry term, for the ``soft'' symmetry term 
$\gamma(t)=0.69\,10^{-3}$ and $\gamma(t)=0.37\,10^{-3}$, with and without 
the contributions from the isovector--like fluctuations, respectively. 
Bottom panel: $\alpha=0.2$,  the value of the 
parameter $\gamma(t)$ is $\gamma(t)=1.62\,10^{-3}$ 
for the ``superstiff'' symmetry term, for the ``soft'' symmetry term 
$\gamma(t)=0.89\,10^{-3}$ and $\gamma(t)=0.56\,10^{-3}$, with and without 
the contributions from the isovector--like fluctuations, respectively.
}
\end{figure}
\par
In Fig.~\ref{fig2} we report the isotopic yields 
of $Z=6$--fragment, calculated according to 
Eqs. (\ref{distrzn}) and (\ref{probzn}) 
for two different values of the asymmetry: 
$\alpha=0.1$ and $\alpha=0.2$. 
The used values of the parameters ${\rm T}=4.5\,{\rm MeV}$, 
$\varrho=0.3\varrho_{eq}$ and $t=125\,{\rm fm}/c$, where $t$ is the time that 
the system spends in the instability region, are compatible with the 
analogous values obtained within the SMF approach of Ref. \cite{Bar02}. 
For the dynamical correlation length we have chosen the value 
$L=1.3\,r_0$. This value corresponds to the effective exponent 
$\tau_{eff}=1.65$ of the power law $Y(Z)=Y_0Z^{-\tau_{eff}}$ 
for fragment distribution \cite{Mat03}.
In the figure we display the results obtained with the ``superstiff'' 
asymmetry term and with the ``soft'' asymmetry term of the 
nucleon--nucleon interaction. 
Moreover we compare also the relative contribution of isoscalar-like
and isovector-like fluctuations to the width. 

In the "superstiff" case isovector--like oscillations are suppressed 
for the considered values of $\varrho$, $T$ and $\alpha$, 
i.e. Eq. (\ref{rate}) has only one pole,  
so the width comes essentially from the dispersion of the chemical effect
in the isoscalar-like fluctuations (diamonds). 
In the ``soft'' case, the full calculation is represented by triangles, while
the result obtained taking into account only the contribution from 
the isoscalar-like modes is 
represented by circles.  Comparing diamonds and circles,  
we observe that the ``superstiff'' 
asymmetry term gives rise to a wider isotopic distribution. 
This is due to the fact that the ``soft'' asymmetry term, at the 
considered density, is more effective to drive fragments 
closer to the average asymmetry value, with respect to an asymmetry 
term with a stronger density dependence.
Indeed we find that, in spite of the competition with Coulomb and surface
effects, the isospin distillation mechanism does not change much with
the wave number $k$, in the ``soft'' case. 
The counterpart in our formalism is that in this case the behaviors 
of the components of the covariance matrix, as 
functions of $k$, are more similar each other 
reducing the width of the isotopic distribution.
However, adding the contributions due to the isovector-like fluctuations, 
the total width obtained in the ``soft'' case (triangles) becomes closer
to the ``superstiff'' results. 
It is also possible to observe that the contribution of the 
isovector-like fluctuations to the full width is more important at 
smaller asymmetry.  This is because isovector-like 
fluctuations become weaker 
when increasing the asymmetry of the matter. 
\par
Figure \ref{fig2} also shows that the width of the isotopic yields 
increases with asymmetry. This corresponds to the general property 
that for more neutron--rich systems the density--density response 
function of neutrons is enhanced with respect to that of protons. 
In addition, we can see that 
the more neutron--rich system ($\alpha=0.2$) produces the more 
neutron--rich isotopes, as expected. 
\par
It is worth to remark that both the overall behavior and the widths 
of the distributions of Fig.~\ref{fig2}  
favourably compare with the corresponding distributions for primary 
fragments calculated within the SMF approach \cite{Liu04}. 
\par
\begin{figure}
\includegraphics{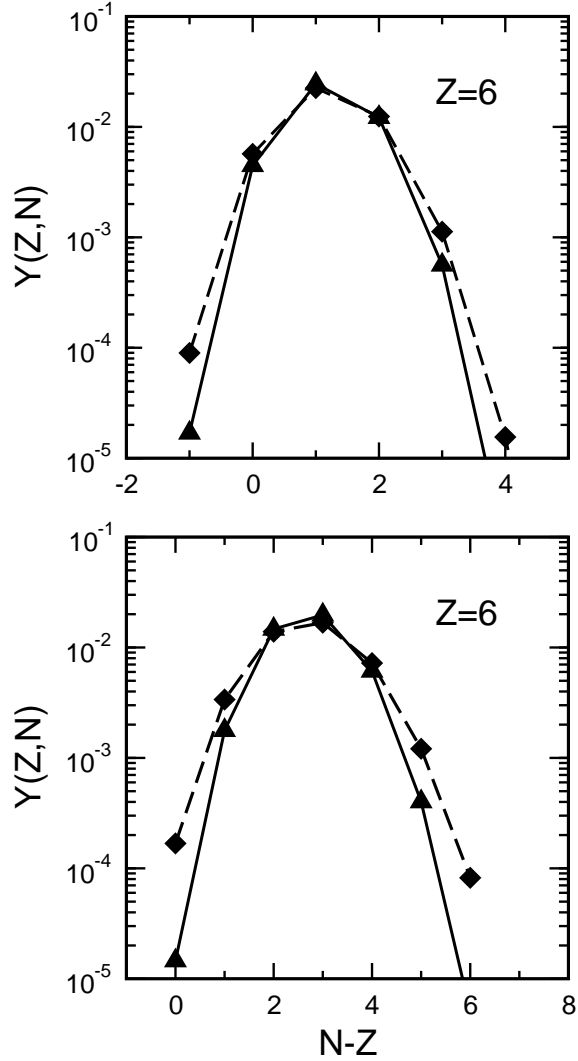}
\caption{\label{fig2_new}Isotopic distributions calculated according to the 
correlation volume prescription (Eq. \ref{distrzn_stat}).  
The values of the parameters and the symbols are the same as in Fig~\ref{fig2}.
}
\end{figure}
In Fig.~\ref{fig2_new} we present isotopic distributions obtained using the 
correlation volume prescription (Eq. \ref{distrzn_stat}), 
with ${\bar A} = 20$. This value corresponds to the average size 
of intermediate mass fragments, 
as obtained in the considered conditions of density and temperature.   
As one can see by comparing Figs.~\ref{fig2} and \ref{fig2_new}, 
results are not very different with the two prescriptions.\par
\begin{figure}
\includegraphics{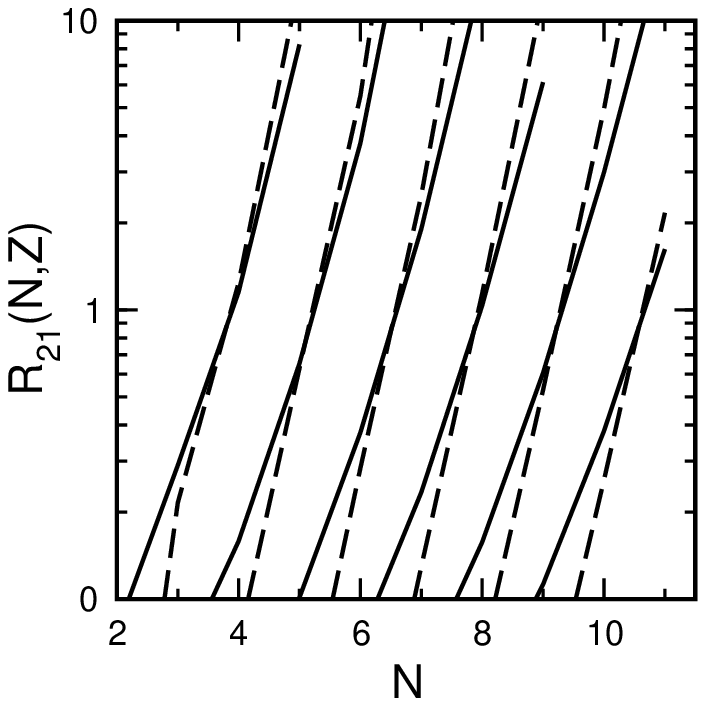}
\caption{\label{fig3}Isotopic ratio 
$R_{21}(N,Z)=Y_{\alpha=0.2}(N,Z)/Y_{\alpha=0.1}(N,Z)$ 
calculated with the ``superstiff'' symmetry term (solid lines ) 
and with the ``soft'' symmetry term (dashed lines). 
Lines correspond to different values of $Z$, $Z=3-8$ from left to 
right. The values of remaining parameters are the same as in Fig~\ref{fig2}.
}
\end{figure}
The ratio between isotopic yields observed in two different reactions, 
$R_{21}(N,Z)=Y_{\alpha_2}(N,Z)/Y_{\alpha_1}(N,Z)$,  
shows a very simple behavior. As a function of $Z$ and $N$, it 
can well be fitted  by an exponential law (the so called 
isoscaling relationship) \cite{Xu00,Tsa01,Tan01,Tsa101}. 
In addition, the isoscaling relationship 
has been reproduced by SMF--model calculations also for the distributions 
of primary fragments \cite{Liu04}. This particular feature 
of the isotopic distributions can represent an effective tool 
to compare isotopic distributions from systems with different 
$N/Z$ ratios. 
\par
\begin{figure}
\includegraphics{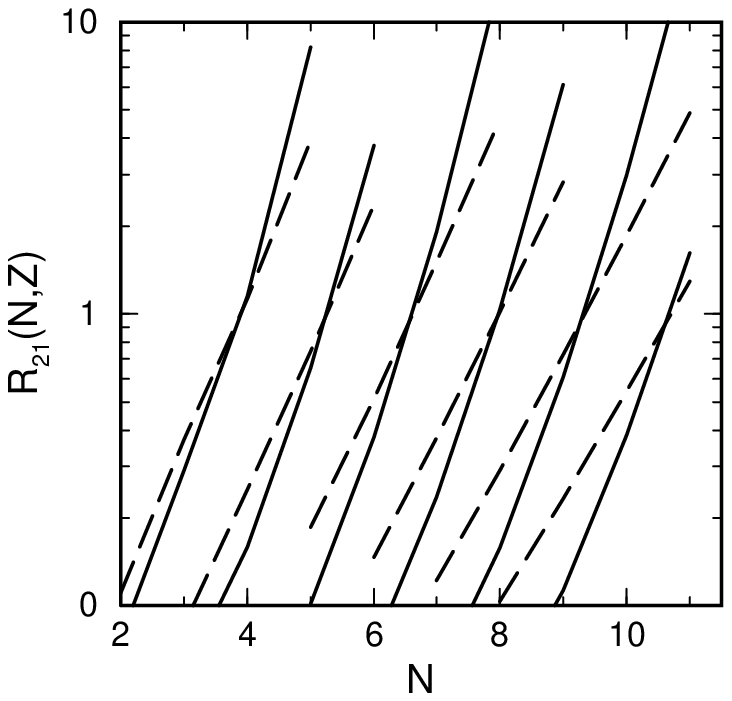}
\caption{\label{fig4}Same as in Fig.~\ref{fig3} but using  
only the ``superstiff'' symmetry term and  
for two different values of the density:   
solid lines correspond to $\varrho=0.3\,\varrho_{eq}$ 
and $t=125\,{\rm fm}/c$, dashed lines correspond to 
$\varrho=0.4\,\varrho_{eq}$ and $t=150\,{\rm fm}/c$. 
}
\end{figure}

The isotopic ratio $R_{21}(N,Z)$ calculated in our approach,
according to Eqs. (\ref{distrzn}) and (\ref{probzn}),  
for two different values of the asymmetry parameter, $\alpha_2=0.2$ and 
$\alpha_1=0.1$, is displayed in Figs.~\ref{fig3} and \ref{fig4}. 
In Fig.~\ref{fig3} we compare the values of $R_{21}(N,Z)$ 
as a function of $N$, obtained 
with the ``superstiff'' symmetry term and with the ``soft'' symmetry term. 
The linear behavior, in logaritmic scale, with the same slope  
for every $Z$ is reproduced in both cases within a satisfying 
approximation. Because of the smaller value of the width 
parameter $\gamma(t)$, the ``soft'' symmetry term 
gives a steeper slope with respect to the ``superstiff'' term. 
The average values of the slope approximatively are $2.2\pm 0.2$ 
and $1.5\pm 0.15$ for the ``soft'' case and the ``superstiff'' case 
respectively. \par
In Fig.~\ref{fig4} the ratio $R_{21}(N,Z)$ is displayed for two values 
of the density of the system at the break--up. In order 
to obtain fluctuations of similar magnitude 
in the two cases, two different times 
the system spends in the instability region are considered.
Nevertheless, a behavior with a steeper slope 
is observed in the more unstable case. This is 
due to a smaller value of $\gamma(t)$ in this case, since, for a given 
charge asymmetry, the response functions of protons and of neutrons 
tend to be more similar with decreasing density. 
\par
\begin{figure}
\includegraphics{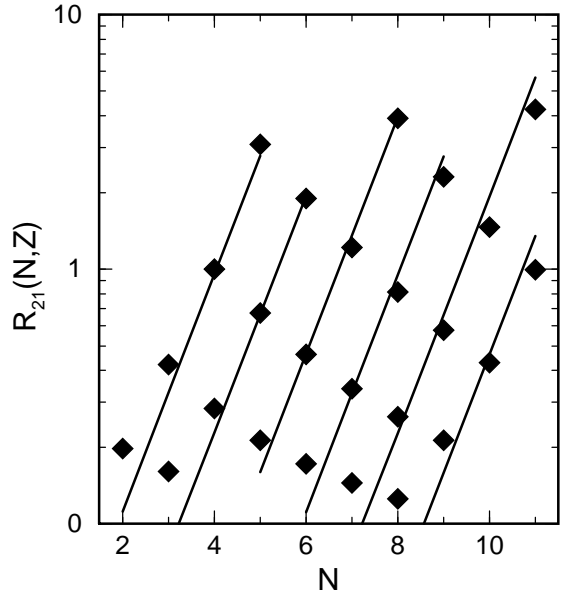}
\caption{\label{fig5}Comparison of calculated isotopic ratio 
$R_{21}(N,Z)=Y_{\alpha=0.195}(N,Z)/Y_{\alpha=0.13}(N,Z)$ (diamonds) 
with the fit for primary fragments of Ref.~\cite{Liu04} (solid lines). 
From left to right $Z=3,4,5,6,7,8$. Calculations are done with the 
``superstiff'' symmetry term. The values of density $\varrho$, 
temperature ${\rm T}$, time $t$, and ratio $L/r_0$ are the same as 
in Fig.~\ref{fig2}. 
}
\end{figure}

We now perform a more quantitative comparison between predictions of our 
approach and results for primary fragments of the SMF--model 
calculations of Ref. \cite{Liu04}. To this purpose we adopt for the  
average asymmetry of fragments the values predicted 
by the SMF model for semicentral collisions of $^{112}{\rm Sn}+^{112}
{\rm Sn }$ and $^{124}{\rm Sn}+^{124}{\rm Sn}$ \cite{Bar02,Liu04}: 
$\alpha_1=0.13$ and $\alpha_2=0.195$, respectively. 
In both the approaches the same ``superstiff'' 
symmetry term for the effective interaction is used. 
Also the values of density 
$\varrho=0.3\varrho_{eq}$, temperature ${\rm T}=4.5\,{\rm MeV}$, and 
time spent at the break--up $t=125\,{\rm fm}/c$ are chosen according to the 
results of SMF--model calculations. Figure~\ref{fig5} shows the isotopic 
ratio $R_{21}(N,Z)$ calculated with our approach 
and the curves obtained in Ref.~\cite{Liu04} by fitting the results 
of the SMF model with an exponential law. We observe a remarkable 
agreement between the results of our nuclear matter calculations 
and the simulations of the SMF--model. 

\section{Conclusions}
In this article we discuss relevant observables of multifragmentation
processes in charge asymmetric nuclear matter, such 
as the isotopic distribution of 
intermediate--mass fragments, as obtained within the spinodal decomposition 
scenario, on the basis of an analytical approach.
Fragmentation happens due to the development of isoscalar-like unstable
modes, i.e. unstable density oscillations with also a chemical component, 
leading to the formation of more symmetric fragments. We 
find that the isotopic distributions are peaked at a value given by the 
average distillation effect, while the width is determined by 
the dispersion of the chemical effect among the relevant unstable modes
and by isovector-like fluctuations present in the matter that undergoes 
spinodal decomposition. 
The size of this dispersion is mostly due to the 
competition between symmetry energy
effects (that favour the formation of symmetric fragments)
and the Coulomb repulsion, that acts against the concentration of protons
in large density domains, expecially for modes with large wavelength.
Clearly the net result of this competion 
also depends on the EOS used. Smaller widths are
obtained with a ``soft'' symmetry energy term. However, 
the contribution due to isovector-like fluctuations is more important 
in the ``soft'' case, indeed in the ``superstiff'' case isovector oscillations
are suppressed. Hence finally the isotopic distributions are quite similar
when using the two parameterizations of the symmetry energy.  

In particular, we find that, when considering two systems with different 
asymmetry, the isotopic (or isotonic) yields obey an approximate
isoscaling, with a slope connected to the difference betwen the asymmetries 
of the two systems and to the differences between the widths 
of the isotopic distributions. 
Hence isoscaling properties can be recovered in a dynamical picture.
We notice that isoscaling has been found in dynamical simulations
of heavy ion collisions, such as 
stochastic mean field \cite{Liu04} and antisymmetrized molecular dynamics  
calculations \cite{Ono}. 

The isoscaling parameters are also connected to the properties
of the symmetry term in the EOS. Indeed we have seen that 
a stiffer behavior of the symmetry energy
term yields larger isotopic widths, leading to smaller values of the 
slope (see Fig. \ref{fig3}).   
However, as reported in Ref. \cite{Bar02}, we also 
observe that in collisions of charge asymmetric systems, 
pre-equilibrium emission 
is less neutron rich when using a stiffer parametrization of the 
symmetry term (thus leading to more asymmetric fragments), with respect to the 
``soft'' case.  Therefore, in the isoscaling analysis, there could be 
a compensation between the average asymmetry of fragments (larger in the
``stiff'' case) and the width of the distribution 
(also larger in the ``stiff'' case).
In fact, for the systems considered in Ref. \cite{Liu04}, 
similar values of the slope are obtained for the two parameterizations
considered for the symmetry energy.

It may also be interesting to notice that the values obtained in our 
calculations are larger than the predictions 
of statistical multifragmentation models, see Ref. \cite{Tsa101}. 
Of course this picture can be modified by the secondary de-excitation 
process, that reduces the asymmetry of fragments and, consequently, 
the slopes deduced from isoscaling.  
Hence the final distributions can be quite different from the primary ones.
A more detailed study, aiming to extract information on the primary 
distributions and on the fragmentation mechanism,
would require the introduction of more sophisticated observables, 
probably based on an event by event analysis, in line with the 
recent investigations of correlations between intermediate--mass 
fragments \cite{Bor02}.

\end{document}